\RequirePackage{ifpdf}
\ifpdf % We are running pdfTeX in pdf mode
\documentclass[pdftex]{sigma}
\else
\documentclass{sigma}
\fi

\usepackage[all]{xy}

\begin{document}
\allowdisplaybreaks

\renewcommand{\PaperNumber}{007}

\FirstPageHeading{shapovalov}

\ShortArticleName{Exact Solutions and Symmetry Operators} % maximum 75 symbols

\ArticleName{Exact Solutions and Symmetry Operators \\ 
for the Nonlocal Gross--Pitaevskii Equation \\
with Quadratic Potential}

\Author{Alexander SHAPOVALOV~$^{\dag\ddag\S}$, Andrey
TRIFONOV~$^{\ddag\S}$ and Alexander LISOK~$^{\S}$}

\AuthorNameForHeading{A. Shapovalov,  A. Trifonov and A. Lisok}

\Address{$^\dag$~Tomsk State University, 36 Lenin Ave., 634050 Tomsk, Russia}
\EmailD{\href{mailto:shpv@phys.tsu.ru}{shpv@phys.tsu.ru}}

\Address{$^\ddag$~Tomsk Polytechnic University, 30 Lenin Ave., 634050 Tomsk, Russia}
\EmailD{\href{mailto:trifonov@mph.phtd.tpu.edu.ru}{trifonov@mph.phtd.tpu.edu.ru}}

\Address{$^\S$~Math. Phys. Laboratory, Tomsk Polytechnic University, 30 Lenin Ave.,  634050 Tomsk,
Russia}
\EmailD{\href{mailto:lisok@mph.phtd.tpu.edu.ru}{lisok@mph.phtd.tpu.edu.ru}}

\ArticleDates{Received July 27, 2005, in final form October 06, 2005; Published online October 17, 2005}

\Abstract{The complex WKB--Maslov method is used to consider an
approach to the semiclassical integrability of the
multidimensional Gross--Pitaevskii equation with an external field and
nonlocal nonlinearity previously developed by the authors.
Although the WKB--Maslov method is approximate in essence, it leads to
exact solution of the Gross--Pitaevskii equation with an external
and a nonlocal quadratic potential. For this equation,
an exact solution of the Cauchy problem is constructed in the class of
trajectory concentrated functions. A~nonli\-near evolution
operator is found in explicit form and symmetry
operators (mapping a solution of the equation into another
solution) are obtained for the equation under consideration.
General constructions are illustrated by examples.}


\Classification{81Q20; 81Q30; 81R30} % e.g. 35A30; 81Q05

\section{Introduction}

Experimental advances in the realization of  Bose--Einstein
condensation (BEC) in  weakly interacting  alkali-metal atomic
gases~\cite{shapovalov:CORNELL} have generated  great interest in
the theoretical study of the BEC. Its states and evolution are
described using the Gross--Pitaevskii equation (GPE)
\cite{shapovalov:PITAEVSKII-1, shapovalov:GROSS} for  the wave
function $\Psi(\vec x,t)$ of the condensate confined by  external
field with potential $V_{\rm ext}(\vec x,t)$ at zero temperature:
\begin{gather}
\label{shapovalov:NLS} \left(-i\hbar\partial_t
+\displaystyle\frac{\hat{\vec p}\,{}^2}{2m} +V_{\rm ext}(\vec
x,t)+\varkappa |\Psi(\vec x,t)|^2 \right)\Psi(\vec x,t)=0.
\end{gather}
Here $\vec x\in\mathbb{R}^n_x$,  $\hat{\vec p}=$
$-i\hbar\partial/\partial \vec x $, $t\in \mathbb{R}^1$, $\partial_t=\partial/\partial t$,
 $|\Psi|^2=\Psi^*\Psi$,
$\Psi^*$ is complex conjugate to $\Psi$, $\varkappa $ is a real nonlinearity parameter,
$|\Psi(\vec x,t)|^2$ is the condensate density, and
$N[\Psi]= \int_{\mathbb R^n} |\Psi(\vec x,t)|^2d\vec x$ is
the number of condensate particles. Following quantum mechanics,
we refer to the solutions of the GPE as states.

Equation (\ref{shapovalov:NLS}) is of nonlinear Schr\"odinger equation type with
the local cubic nonlinearity $\varkappa |\Psi(\vec x,t)|^2$
rep\-re\-senting the boson interaction in the mean field approximation.
Besides the BEC,  equation (\ref{shapovalov:NLS}) describes a wide
spectrum of nonlinear phenomena such as instability of water waves,
nonlinear modulation of collisionless plasma waves,
optical pulse propagation in nonlinear media and others.
In all these cases, space localized
soliton-like solutions are of principal interest.
However, the  wave packets described by equation (\ref{shapovalov:NLS})  with $V_{\rm ext}=0$
 in  multidimensional space  ($n>1$) with
focusing  nonlinearity ($\varkappa<0$) are known to collapse, i.e.
the situation where the wave amplitude increases extremely  and
becomes singular within  a finite time or propagation distance
(see, for example,  \cite{shapovalov:KIVSHAR} for reviews). To
eliminate the collapse, which is  considered as an artifact of the
theory, one has to consider some effects  that would render
collapse impossible. The nonlocal form of nonlinearity is
significant since it can, basically,  eliminate collapse in all
physical dimensions ($n=2,3$) \cite{shapovalov:BANG}. Meanwhile,
in the derivation of the GPE (\ref{shapovalov:NLS}), a nonlocal
nonlinearity term $\int_{\mathbb {R}^n}V(\vec x,\vec x')$
$|\Psi(\vec x',t)|^2d \vec x' $ $\Psi(\vec x,t)$ arises, and it is
reduced to $\varkappa |\Psi(\vec x,t)|^2 \Psi(\vec x,t)$,
$\varkappa =\int V(\vec x)d \vec x$ if the potential $V(\vec x,
\vec x')=V(\vec x-\vec x')$ is short-range
\cite{shapovalov:PITAEVSKII-1}.

In view of this consideration, of fundamental interest is the
study of both the properties and the localized solutions of the
nonlocal Gross--Pitaevskii equation. The integrability of the GPE
is a nontrivial problem. In the one-dimensional case ($n=1$) with
$V_{\rm ext}=0$,  equation (\ref{shapovalov:NLS}) (called the nonlinear
Schr\"odinger equation (NLSE)) is known to be exactly integrable
by the Inverse Scattering Trans\-form (IST) method
\cite{shapovalov:ZAKHAROV-1, shapovalov:ZAKHAROV-2}. This is the
only IST  integrable case since with $V_{\rm ext}\ne 0$ the IST fails
even in the one-dimensional case. The same is true for both the
nonlocal GPE in all dimensions and the local GPE
(\ref{shapovalov:NLS}) in a multidimensional case ($n>1$). Direct
application of symmetry analysis
\cite{shapovalov:OVS,shapovalov:IBRAGIM,shapovalov:OLVER,
shapovalov:FUSCH-SS,shapovalov:FUSCH-N} to the nonlocal GPE is
also hampered by presence of the nonlocal term and external
potential in the equation.

In \cite{shapovalov:BTS1,shapovalov:BTS2,shapovalov:KIEV03,shapovalov:LTS}
a {\it semiclassical integrability} approach was
developed for a generalized nonlocal GPE named there a Hartree type equation:
\begin{gather}
\big\lbrace -i\hbar\partial _t +\hat {\mathcal
H}_{\varkappa}(t)\big\rbrace\Psi(\vec x,t)=
\big\lbrace -i\hbar\partial _t +\hat {\mathcal
H}(t)+\varkappa\hat V(t,\Psi (t))\big\rbrace\Psi(\vec x,t) =0, \label{shapovalov:GPE}\\
\Psi(\vec x,t)\in L_2({\mathbb R}^n_x),\qquad  \hat V(t,\Psi(t) )=
\displaystyle\int_{{\mathbb R}^n} d\vec y\,\Psi^*(\vec y ,t)
V(\hat z,\hat w,t)\Psi (\vec y ,t). \label{shapovalov:VTE-2}
\end{gather}
Here the linear operators $\hat{\mathcal H}(t)={\mathcal H}(\hat z, t)$ and
$V(\hat z,\hat w,t)$  are Weyl-ordered functions
\cite{shapovalov:KARASEVMASLOV} of time~$t$ and
of noncommuting operators
\[ \hat z=( \hat{\vec p},\vec x)=(-i\hbar\displaystyle\partial/{\partial\vec x}, \vec x), \qquad
\hat w=( -i\hbar\displaystyle\partial/{\partial\vec y}, \vec y), \qquad
\vec x,\vec y\in {\mathbb R}^n, \]
with commutators
\begin{gather}
\left[ \hat z_k,\hat z_j \right]_- =\left[ \hat
w_k,\hat w_j \right]_- = {i}\hbar J_{kj},
\qquad  \left[ \hat z_k,\hat w_j \right]_- = 0,\qquad
k,j=\overline {1, 2n}, \label{shapovalov:COMM-1}
\end{gather}
$J =\|J_{kj}\|_{2n\times 2n}$ is a identity symplectic matrix
$ J=\left(\begin{array}{cc}0&-{\mathbb I}\\
{\mathbb I}& 0 \end{array} \right)_{2n\times 2n}$, ${\mathbb
I}={\mathbb I}_{n\times n}$ is an identity $(n\times n)$-matrix. This
approach is based on the WKB--Maslov complex germ theory
\cite{shapovalov:MASLOV-1,shapovalov:BEL-DOB} and gives a formal
solution of the Cauchy problem, asymptotic in formal small
parameter $\hbar$ ($\hbar\to 0$)  accurate to $O\big(\hbar^{N/2}\big)$,
where $N$ is any natural number. The Cauchy problem was considered
in the ${\mathcal P}_\hbar^t$ class
 of  trajectory concentrated functions (TCFs) introduced in
\cite{shapovalov:BTS1,shapovalov:BTS2}. Being  approximate one in
essence, the semiclassical approach results in some cases in  {\it
exact} solutions.

In the present work we construct an exact solution of the Cauchy
problem in  the ${\mathcal P}_\hbar^t$ class   for  equation
(\ref{shapovalov:GPE}) with the linear operators ${\mathcal H}
(\hat z, t)$ and $ V(\hat z,\hat w,t)$ being quadratic in $\hat z$,
$\hat w$:
\begin{gather}
{\mathcal H} (\hat z, t)= \frac{1}{2}\langle\hat z,{\mathcal H}_{zz}(t)\hat z\rangle+
\langle {\mathcal H}_z(t),\hat z\rangle, \label{shapovalov:QUAD-1}\\
 V(\hat z,\hat w,t) =
\frac{1}{2}\langle\hat z,W_{zz}(t)\hat
z\rangle+\langle\hat z,W_{zw}(t)\hat w\rangle+\frac{1}{2}\langle\hat w,W_{ww}(t)\hat
w\rangle . \label{shapovalov:QUAD-2}
\end{gather}
Here,
${\mathcal H}_{zz}(t)$, $W_{zz}(t)$, $W_{zw}(t)$,  $W_{ww}(t)$ are $2n\times 2n$ matrices,
${\mathcal H}_z(t)$ is a $2n$-vector; $\langle \cdot,\cdot\rangle$ is an Euclidean scalar product of vectors:
$\langle\vec p,\vec x\rangle=\sum\limits^n_{j=1}p_jx_j$;\,\, $
\vec p,\vec x\in {\mathbb R}^n$, $ \langle z,w\rangle=\sum\limits^{2n}_{j=1}z_jw_j$,
$z,w\in {\mathbb R}^{2n}$.
By solving the Cauchy problem, we obtain a nonlinear evolution operator
in explicit form. With  the evolution operator obtained, we formulate
a  nonlinear superposition principle for the solutions of the nonlocal
GPE in the class of TCFs.  Also, we give symmetry operators
in general form that map each solution of the GPE into another solution.
The general constructions are illustrated by examples.

\section{Cauchy problem }

Here we give a brief account of the solution of the Cauchy problem for
equations (\ref{shapovalov:GPE}), (\ref{shapovalov:VTE-2}),
(\ref{shapovalov:QUAD-1}), and~(\ref{shapovalov:QUAD-2}),
 following \cite{shapovalov:BTS1, shapovalov:BTS2,shapovalov:KIEV03}.

\subsection*{The class of trajectory concentrated functions}

To set the Cauchy problem for equations (\ref{shapovalov:GPE}),
(\ref{shapovalov:VTE-2}), (\ref{shapovalov:QUAD-1}), and
(\ref{shapovalov:QUAD-2}), following \cite{shapovalov:BTS1,
shapovalov:BTS2,shapovalov:KIEV03}, we  define a class of  functions ${\mathcal
P}_\hbar^t$ via its generic element $\Phi (\vec x,t,\hbar)$:
\begin{gather}
{\mathcal P}_\hbar^t=  \left\lbrace\Phi :\Phi (\vec
x,t,\hbar)= \varphi\left(\frac{\Delta\vec
x}{\sqrt{\hbar}},t,\hbar\right)
\exp\left[{\frac{i}{\hbar}(S(t,\hbar)+ \langle\vec
P(t,\hbar),\Delta\vec x \rangle )}\right]\right\rbrace.
\label{shapovalov:TCF}
\end{gather}
Here, the  function $\varphi(\vec\xi ,t,\hbar)$ belongs to the
Schwartz space ${\mathcal S}({\mathbb R}^n)$ with respect to $\vec\xi\in{{\mathbb
R}}^n$, smoothly depends on $t$, and is regular in $\sqrt\hbar$
for $\hbar\to 0$, $\Delta\vec x= \vec x-\vec X(t,\hbar)$. The
real function $S(t,\hbar)$ and the $2n$-dimensional vector
function $Z(t,\hbar)=(\vec P(t,\hbar),\vec X(t,\hbar))$ define the
${\mathcal P}_\hbar^t$ class, regularly depend on  $\sqrt\hbar$ in
the  neighborhood of  $\hbar=0$,  and  are to be determined. The
functions of the ${\mathcal P}_\hbar^t$ class  are normalizable
with  respect to the norm $\|\Phi (t)\|^2={\langle  \Phi (t) |\Phi
(t) \rangle }$, where
\begin{gather}
\label{shapovalov:NORM}
{\langle  \Psi (t) |\Phi (t) \rangle
}=\int_{\mathbb R^n} d \vec x {\Psi}^* (\vec x,t,\hbar)\Phi
(\vec x,t,\hbar)
\end{gather}
is a scalar product in
the  space $L_2({\mathbb R}^n_x)$.
At any time $t\in{\mathbb R}^1$ the function  $\Phi (\vec x,t,\hbar)\in {\mathcal P}_\hbar^t$
is localized in the limit $\hbar\to 0$ in the neighborhood of a point
of the phase curve $z=Z(t,0)$. For this reason we call
${\mathcal P}_\hbar^t$  {\it the class of trajectory concentrated functions}.

For $t=0$ the ${\mathcal P}_\hbar^t$ class   transforms to  the ${\mathcal P}_\hbar^0$ class
 of functions
\begin{gather}
\psi(\vec x,\hbar)=\exp\left\lbrace\frac i\hbar[S(0,\hbar)+\langle\vec P_0(\hbar),
(\vec x-\vec X_0(\hbar))\rangle]\right\rbrace\varphi_0\left(\frac{\vec x-\vec
X_0(\hbar)} {\sqrt\hbar},\hbar\right),\nonumber\\
\varphi_0(\vec\xi,\hbar)\in{\mathcal S}({\mathbb R}^n_\xi),\label{shapovalov:TCF0}
\end{gather}
where  $Z_0(\hbar)=(\vec P_0(\hbar),\vec X_0(\hbar))$ is a point
of the phase space ${\mathbb R}^{2n}_{px}$, and the constant
$S_0(\hbar)$ can be omitted  without loss of generality. The Cauchy
problem is formulated for equations (\ref{shapovalov:GPE}),
(\ref{shapovalov:VTE-2}), (\ref{shapovalov:QUAD-1}), and
(\ref{shapovalov:QUAD-2}) in the ${\mathcal P}_\hbar^t$ class  of
trajectory concentrated functions as
\begin{gather}
\label{shapovalov:CAUCHY}
\Psi(\vec x,t,\hbar)|_{t=0}=\psi(\vec x,\hbar), \qquad
\psi\in {\mathcal P}^0_\hbar.
\end{gather}

\subsection*{Hamilton--Ehrenfest system}

To solve the Cauchy problem  (\ref{shapovalov:GPE}),
(\ref{shapovalov:VTE-2}), (\ref{shapovalov:QUAD-1}),
(\ref{shapovalov:QUAD-2}), and (\ref{shapovalov:CAUCHY}), we first
obtain the Hamilton--Ehrenfest system (HES) of equations in the
moments of the solution $\Psi(\vec x,t,\hbar)\in {\mathcal
P}^t_\hbar$ of equation (\ref{shapovalov:GPE}).

The operators ${\mathcal H} (\hat z,t)$, equation
(\ref{shapovalov:QUAD-1}), and  $V(\hat z,\hat w ,t)$, equation
(\ref{shapovalov:QUAD-2}), are  self-adjoint, respectively,  to
the scalar product (\ref{shapovalov:NORM}) and to scalar product
\[ \langle\Psi(t)|\Phi(t)\rangle _{{\mathbb R}^{2n}}
= \int_{{\mathbb R}^{2n}} d\vec xd\vec y {\Psi}^* (\vec x,\vec y,t,\hbar)
\Phi(\vec x,\vec y,t,\hbar)
\]
in the space  $L_2({\mathbb R}^{2n}_{xy})$.

Define  the mean value $\langle  \hat A \rangle$  for a  linear operator $\hat A(t)=A(\hat z,t)$
and a state $\Psi(\vec x,t,\hbar)$ as
\begin{gather}
\label{shapovalov:MEAN}
\langle  \hat A \rangle= 
\frac{1}{\|\Psi (t) \|^2}\langle\Psi(t)| \hat A| \Psi(t)
\rangle=A_\Psi(t,\hbar).
\end{gather}
From (\ref{shapovalov:GPE}), (\ref{shapovalov:VTE-2}) we have
\begin{gather} 
 \frac{d}{dt}\langle \hat A(t)\rangle =\Bigl\langle \frac{\partial\hat A(t)} {\partial
t}\Bigr\rangle +\frac{i}{\hbar}\langle [\hat{\mathcal H},\hat A (t)]_-\rangle \nonumber\\
\phantom{\frac{d}{dt}\langle \hat A(t)\rangle =}{} 
+ \frac{i\varkappa}{\hbar}\Bigl\langle\int d\vec y\,\Psi^*(\vec y,t,\hbar) [
V(\hat z,\hat w,t),\hat A(t)]_-\Psi(\vec y,t,\hbar)\Bigl\rangle, \label{shapovalov:EHREN}
\end{gather}
where $[\hat A,\hat B]_-=\hat A\hat B-\hat B\hat A$ is
the commutator of the operators $\hat A$ and $\hat B$.
We
refer to equation (\ref{shapovalov:EHREN}) as the {\em Ehrenfest equation for
the operator $\hat A$ and function}\ $\Psi(\vec x,t,\hbar)$ as in
quantum mechanics~\cite{shapovalov:EHRENFEST}.

For  $\hat A=1$,  equation (\ref{shapovalov:EHREN}) gives
$\|\Psi(t)\|^2=\|\Psi(0)\|^2=\|\Psi\|^2$.
This implies that the norm of a solution of equations (\ref{shapovalov:GPE}) and (\ref{shapovalov:VTE-2})
does not depend on time, and we can use the parameter $\tilde\varkappa=\varkappa\|\Psi\|^2$
instead of $\varkappa$ in (\ref{shapovalov:GPE}).

Assume that
\begin{gather}
\Delta_{\Psi\alpha}(t,\hbar)=\frac{\langle\Psi(t)|\lbrace\Delta{\hat z}\rbrace^\alpha|\Psi(t)\rangle}
{\|\Psi\|^2},
\qquad \alpha\in {\mathbb Z}_+^{2n}, \label{shapovalov:MOMENTS-g}
\end{gather}
are the
moments
of order $|\alpha|$ of  the function $\Psi(\vec x,t)$ centered
with respect to $z_\Psi(t,\hbar)=$ $(\vec p_\Psi(t,\hbar)$, $\vec
x_\Psi(t,\hbar))$. Here  $\lbrace\Delta{\hat z}\rbrace^\alpha$ is
an operator with a Weyl symbol  $(\Delta z_\Psi)^\alpha$,
\[
\Delta z_\Psi=z-z_\Psi(t,\hbar)=(\Delta\vec p_\Psi,\Delta\vec x_\Psi),
\qquad \Delta\vec p_\Psi=\vec p-\vec p_\Psi(t,\hbar),
\qquad \Delta\vec x_\Psi=\vec x-\vec x_\Psi(t,\hbar) .\]

Along with (\ref{shapovalov:MOMENTS-g}) we use the following notation for
the variances of coordinates, momenta, and correlations:
\begin{gather}
\Delta_{{\Psi}2}(t,\hbar)=
\|\Delta_{\Psi jk}(t,\hbar)\|_{2n\times 2n}=
 \frac{1}{2\|\Psi\|^2}\|\langle\Psi(t)| \lbrace
\Delta\hat z_j\Delta\hat z_k+ \Delta
\hat z_k\Delta\hat z_j\rbrace|\Psi(t)\rangle\|_{2n\times 2n} \nonumber\\
\phantom{\Delta_{{\Psi}2}(t,\hbar)}{}
=\left(\begin{array}{ll}{\sigma}_{pp}(t,\hbar) & {\sigma}_{ px}(t,\hbar)\\
{\sigma}_{xp}(t,\hbar) & {\sigma}_{xx}(t,\hbar)
\end{array}\right),
\nonumber\\
  {\sigma}_{xp}(t,\hbar)= \frac{1}{2}\|\langle \lbrace \Delta
x_j\Delta\hat p_k+ \Delta\hat p_k\Delta x_j\rbrace\rangle\|_{n\times n}, \nonumber\\
 {\sigma}_{xx}(t,\hbar)=
\|\langle\Delta x_j\Delta x_k
\rangle\|_{n\times n}, \qquad {\sigma}_{pp}(t,\hbar)=
\|\langle
\Delta\hat p_j\Delta
\hat p_k \rangle\|_{n\times n}. \label{shapovalov:SIGMA}
\end{gather}
The Ehrenfest equations (\ref{shapovalov:EHREN}) in  mean values
can be obtained for the operators $\hat z_j$, $\lbrace\Delta\hat
z\rbrace^\alpha$ and trajectory concentrated functions
(\ref{shapovalov:TCF}) (see \cite{shapovalov:BTS1,shapovalov:BTS2}
for details). However,  for solving the equations under
consideration, equations (\ref{shapovalov:GPE}),
(\ref{shapovalov:VTE-2}), (\ref{shapovalov:QUAD-1}), and
(\ref{shapovalov:QUAD-2}), we need only equations for the first-order
and second-order moments
\begin{gather}
  \dot z_{\Psi}=J\lbrace{\mathcal H}_z(t)+[{\mathcal H}_{zz}(t)+
\tilde\varkappa(W_{zz}(t)+W_{zw}(t))]z_{\Psi}\rbrace ,\nonumber\\
 \dot\Delta_{\Psi 2}=J[{\mathcal H}_{zz}(t)+\tilde\varkappa W_{zz}(t)]
 \Delta_{\Psi 2}-\Delta_{\Psi 2} [{\mathcal H}_{zz}(t)+\tilde\varkappa W_{zz}(t)]J.\label{shapovalov:HES}
\end{gather}
We call the system (\ref{shapovalov:HES}) {\it the Hamilton--Ehrenfest
system} (HES) of the second order  for equations
(\ref{shapovalov:VTE-2}), (\ref{shapovalov:QUAD-1}), and
(\ref{shapovalov:QUAD-2}). Following
\cite{shapovalov:BTS1,shapovalov:BTS2}, we equate the functional
vector-parameter $Z(t,\hbar)$ of the ${\mathcal P}_\hbar^t$ class
of TCFs (\ref{shapovalov:TCF}) with $z_\Psi(t,\hbar)$, i.e. $\vec
p_\Psi(t,\hbar)=\vec P(t,\hbar)$,
 $\vec x_\Psi(t,\hbar)=\vec X(t,\hbar)$.
This relates  the ansatz (\ref{shapovalov:TCF}) to an exact solution of equation
(\ref{shapovalov:GPE}).

Consider a phase space ${\mathcal M}^N$, $\dim {\mathcal
M}^N=N=3n+2n^2$, of points ${\mathfrak g}\in {\mathcal M}^N$ with
coordinates
\begin{gather*}
{\mathfrak g}=(z, \Delta)^\intercal, \qquad z\in {\mathbb R}^{2n},
\qquad z=(\vec p,\vec x)^\intercal,  \\
\Delta=(\Delta_{ij})^\intercal, \qquad \Delta_{ij}=\Delta_{ji},
\qquad \Delta\in{\mathbb R}^{n+2n^2}, \qquad  i,j =1,\dots ,2n.
\end{gather*}
Here $A^\intercal$ is a matrix transposed to $A$. The coordinates of
${\mathfrak g}\in {\mathcal M}^N$ are written as matrix columns.
The HES (\ref{shapovalov:HES}) can be considered  a dynamic system
in ${\mathcal M}^N$:
\begin{gather}
\dot z=J\lbrace{\mathcal H}_z(t)+[{\mathcal H}_{zz}(t)+
\tilde\varkappa(W_{zz}(t)+W_{zw}(t))]z \rbrace ,\label{shapovalov:HES-ABST} \\
\dot\Delta=J[{\mathcal H}_{zz}(t)+\tilde\varkappa W_{zz}(t)]
\Delta-\Delta [{\mathcal H}_{zz}(t)+\tilde\varkappa W_{zz}(t)]J. \label{shapovalov:HES-ABST1}
\end{gather}
%%%%%%%%%%%%%%%%%%%%
With the substitution
\[\Delta(t)= A(t)\Delta(0) A^+(t), \]
equation (\ref{shapovalov:HES-ABST1})
is rewritten in equivalent form:
\begin{gather}
\dot A=J[{\mathcal H}_{zz}(t)+\tilde\varkappa W_{zz}(t)]A, \qquad A(0)={\mathbb  I}.  \label{shapovalov:VARIATS}
\end{gather}
We call equation (\ref{shapovalov:VARIATS}) {\it a  system in variations}.

Denote by ${\mathfrak g}(t,{\mathfrak C})$
the general solution  of equations (\ref{shapovalov:HES-ABST}),
(\ref{shapovalov:HES-ABST1}):
\begin{gather}
{\mathfrak g}(t,{\mathfrak C})=\big(\vec P(t,\hbar,{\mathfrak C}),\vec
X(t,\hbar,{\mathfrak C}),\Delta_{11}(t,\hbar,{\mathfrak C}), \Delta_{12}(t,\hbar,{\mathfrak C}), \ldots ,
\Delta_{2n 2n}(t,\hbar,{\mathfrak C})\big)^\intercal \label{shapovalov:HES-SOL}
\end{gather}
and by $\hat{\mathfrak g}$ --- the  operator column
\begin{gather}
\hat{\mathfrak g}=\big(\hat{\vec p}, \vec{ x},(\Delta\hat p_1)^2,\Delta\hat p_1\Delta\hat p_2,\ldots,(\Delta
x_n)^2\big)^\intercal.\label{shapovalov:HES-G}
\end{gather}
Here
\begin{gather}
\label{shapovalov:CONS-0}
{\mathfrak C}=\big(C_1,\ldots,C_N\big)^\intercal\in {\mathbb R}^{3n+2n^2}
\end{gather}
are arbitrary constants.
Given constants ${\mathfrak C}$,  equation (\ref{shapovalov:HES-SOL}) describes
a trajectory of a point in the phase space ${\mathcal M}^N$.
\begin{lemma}\label{shapovalov:THEOR1}
Let  $\Psi(\vec x,t)$ be a partial solution of the GPE
\eqref{shapovalov:GPE}, \eqref{shapovalov:VTE-2},
\eqref{shapovalov:QUAD-1}, \eqref{shapovalov:QUAD-2}
with an initial solution  $\Psi(\vec x,t)\big|_{t=0}=
\psi(\vec x)$. Define the constants ${\mathfrak C}(\Psi(t))$ by the condition
\begin{gather}
{\mathfrak g}(t,{\mathfrak C})= \frac{1}{\|\Psi \|^2}
\langle \Psi(t)|\hat{\mathfrak g}| \Psi(t)\rangle, \label{shapovalov:CONS-1}
\end{gather}
and the constants ${\mathfrak C}(\psi)$ by the condition
\begin{gather}
{\mathfrak g}(0,{\mathfrak C})= \frac{1}{\|\psi \|^2}\langle\psi|\hat{\mathfrak g}|
\psi\rangle.\label{shapovalov:CONS-2}
\end{gather}
Then,
\begin{gather}
{\mathfrak C}(\Psi(t))={\mathfrak C}(\psi),
\label{shapovalov:CONS-2a}
\end{gather}
i.e., ${\mathfrak C}(\Psi(t))$ are integrals of motion for  equations
\eqref{shapovalov:GPE}, \eqref{shapovalov:VTE-2}, 
\eqref{shapovalov:QUAD-1}, and \eqref{shapovalov:QUAD-2}.
\end{lemma}

\begin{proof} By construction, the vector
\begin{gather}
{\mathfrak g}(t)= \frac{1}{\|\Psi \|^2}\langle\Psi(t)|\hat{\mathfrak g}|
\Psi(t)\rangle={\mathfrak g}(t,{\mathfrak C}(\Psi(t)))\label{shapovalov:CONS-3}
\end{gather}
is a partial solution of the system (\ref{shapovalov:HES-ABST}),
(\ref{shapovalov:HES-ABST1}),
which  coincides with ${\mathfrak g}(t,{\mathfrak C}(\psi))$
at the initial time  $t=0$. In view  of uniqueness of the solution of the Cauchy problem
for the system (\ref{shapovalov:HES-ABST}) and~(\ref{shapovalov:HES-ABST1}), the equality
\begin{gather}
{\mathfrak g}(t,{\mathfrak C}(\psi))=
 {\mathfrak g}(t,{\mathfrak C}(\Psi(t))),\label{shapovalov:CONS-4}
\end{gather}
is valid. The proof is complete.
\end{proof}

\subsection*{Linear associated Schr\"odinger equation}

Let us substitute (\ref{shapovalov:QUAD-1}), (\ref{shapovalov:QUAD-2})
into (\ref{shapovalov:GPE}), (\ref{shapovalov:VTE-2}) and replace
the operators  $\hat z=(\hat{\vec p}, {\vec x})$ and
$\hat w$
by
$\Delta \hat z=\hat z-z(t,\hbar,{\mathfrak C})=(\Delta\hat{\vec p}, \Delta{\vec x})=
(\hat{\vec p}-\vec P(t,\hbar,{\mathfrak C})$, ${\vec x}-\vec X(t,\hbar,{\mathfrak C}))$ and
$\Delta \hat w=\hat w-z(t,\hbar,{\mathfrak C})$.
Then we have
\begin{gather}
 \lbrace -i\hbar\partial_t
+\hat{\mathfrak H}(t,\Psi(t))\rbrace\Psi=0,  \label{shapovalov:GPE-2a}\\
\hat{\mathfrak H}(t,\Psi(t))= 
{\mathfrak H}(t,\Psi(t))+\langle{\mathfrak H}_z(t,\Psi(t)),\Delta\hat z\rangle
+ \frac 12\langle\Delta\hat
z,{\mathfrak H}_{zz}(t,\Psi(t))\Delta\hat z\rangle, \label{shapovalov:GPE-2}\\
{\mathfrak H}(t,\Psi(t))=  \frac
12\langle z_\Psi(t,\hbar),[{\mathcal H}_{zz}(t)+
\tilde\varkappa(W_{zz}(t)+2W_{zw}(t)+W_{ww}(t))]z_\Psi(t,\hbar)\rangle\nonumber\\ 
\phantom{{\mathfrak H}(t,\Psi(t))=}{}+\langle {\mathcal
H}_z(t),z_\Psi(t,\hbar)\rangle+  \frac
12\tilde\varkappa \mbox{Sp}(W_{ww}(t)\Delta_{\Psi 2}),
\nonumber\\
{\mathfrak H}_z(t,\Psi(t))=
{\mathcal H}_z(t)+ ({\mathcal H}_{zz}(t)
+\tilde\varkappa W_{zz}(t)+\tilde\varkappa W_{zw}(t))z_\Psi(t,\hbar),\\
{\mathfrak H}_{zz}(t,\Psi(t))= 
{\mathcal H}_{zz}(t)+\tilde\varkappa W_{zz}(t).\label{shapovalov:GPE-2b}
\end{gather}
Let us replace the mean values
$z_\Psi(t,\hbar)$, $\Delta_\psi(t,\hbar)$
in the nonlinear GPE (\ref{shapovalov:GPE-2})
by the respective terms of the general solution
(\ref{shapovalov:HES-SOL}) of the system (\ref{shapovalov:HES-ABST}),
(\ref{shapovalov:HES-ABST1}).
As a result,  we obtain a linear equation:
\begin{gather}
\lbrace -i\hbar\partial_t +
\hat{\mathfrak H}(t,{\mathfrak g}(t,{\mathfrak C}))\rbrace
\Phi(\vec x,t, \hbar, {\mathfrak C})=0,\label{shapovalov:LASE}\\[6 pt]
\hat{\mathfrak H}(t,{\mathfrak g}(t,{\mathfrak C}))= 
{\mathfrak H}(t,\hbar,{\mathfrak C})+\langle{\mathfrak H}_z(t,\hbar,{\mathfrak C}),\Delta\hat z\rangle
+ \frac 12\langle\Delta\hat
z,{\mathfrak H}_{zz}(t)\Delta\hat z\rangle, \label{shapovalov:LASE-OPR}\\
 {\mathfrak H}(t,\hbar,{\mathfrak C})=
 \frac 12\langle z(t,\hbar,{\mathfrak C}),[{\mathcal H}_{zz}(t)+
\tilde\varkappa(W_{zz}(t)+2W_{zw}(t)+W_{ww}(t))]z(t,\hbar,{\mathfrak C})\rangle \nonumber\\ 
\phantom{{\mathfrak H}(t,\hbar,{\mathfrak C})=}{}+\langle {\mathcal H}_z(t),z(t,\hbar,{\mathfrak C})\rangle+
 \frac 12\tilde\varkappa \mbox{Sp}(W_{zz}(t)\Delta(t,\hbar,{\mathfrak C})),
\nonumber\\
{\mathfrak H}_z(t,\hbar,{\mathfrak C})={\mathcal H}_z(t)+ ({\mathcal H}_{zz}(t)
+\tilde\varkappa W_{zz}(t)+\tilde\varkappa W_{zw}(t))z(t,\hbar,{\mathfrak C}),\nonumber\\
{\mathfrak H}_{zz}(t)= {\mathcal H}_{zz}(t)+\tilde\varkappa W_{zz}(t),
\qquad z(t,\hbar,{\mathfrak C})=(\vec P(t,\hbar,{\mathfrak C}),\vec X(t,\hbar,{\mathfrak C})).\nonumber
\end{gather}
We call equation (\ref{shapovalov:LASE})  {\em the linear associated  Schr\"odinger
equation} (LASE) for the nonlinear Gross--Pitaevskii equation~(\ref{shapovalov:GPE-2}).
More precisely,  equation (\ref{shapovalov:LASE}) is to be considered as a family of
equations parametrized by the constants ${\mathfrak C}$ of the form
(\ref{shapovalov:CONS-0}).
Each element of the family (\ref{shapovalov:LASE}) is 
a~linear Schr\"odinger equation with a quadratic Hamiltonian
with respect to operators of coordinates and momenta. Such an equation is well known
to be solvable  in  explicit form (see, for example,~\cite{shapovalov:MANKO,shapovalov:PEREL}).
In particular,  partial solutions can be found as
Gaussian wave packets and a  Fock basis of solutions, and  Green function
can be constructed.

\subsection*{LASE solutions and GPE solutions}

Consider a relationship between the solutions of the LASE and  GPE.
Let  $\Phi (\vec x, t, \hbar, {\mathfrak C})$ be
the solution of the Cauchy problem for the LASE (\ref{shapovalov:LASE})
\begin{gather}
\label{shapovalov:LASE-C}
\Phi (\vec x, t, \hbar, {\mathfrak C})|_{t=0}=\psi(\vec x, \hbar),
 \qquad \psi\in {\mathcal P}^0_\hbar.
\end{gather}
The function $\Phi (\vec x, t, \hbar, {\mathfrak C})$ depends on
arbitrary parameters  ${\mathfrak C}$ which appear in the LASE~(\ref{shapovalov:LASE}).
Let ${\mathfrak C}$ are subject to the condition~(\ref{shapovalov:CONS-2}) and are functionals
${\mathfrak C}(\psi)$.
\begin{theorem}\label{shapovalov:THEOR2}
The solution of the Cauchy problem
\eqref{shapovalov:CAUCHY} for the GPE \eqref{shapovalov:GPE} is
\begin{gather}
\label{shapovalov:CAUCHY-SOL}
\Psi(\vec x,t,\hbar)=\Phi (\vec x, t, \hbar, {\mathfrak C}(\psi)).
\end{gather}
\end{theorem}

\begin{proof}
The function $\Phi (\vec x, t, \hbar, {\mathfrak
C}(\psi))$ satisfies equation (\ref{shapovalov:LASE}) for arbitrary
${\mathfrak C}$ and also for ${\mathfrak C}=$ ${\mathfrak
C}(\psi)$. According to  equation (\ref{shapovalov:CONS-2a}), in  
Lemma~\ref{shapovalov:THEOR1} we do not violate the equality in
(\ref{shapovalov:LASE}) if  we  replace  ${\mathfrak C}(\psi)$  by
${\mathfrak C}(\Psi(t))$. In view of equations
(\ref{shapovalov:CONS-3}), (\ref{shapovalov:CONS-4}), one can see
that $\Phi (\vec x, t, \hbar, {\mathfrak C}(\psi))= \Phi (\vec
x, t, \hbar, {\mathfrak C}(\Psi(t)))$ and the operator
$\hat{\mathfrak H}(t,{\mathfrak g}(t,{\mathfrak C}))$ in
(\ref{shapovalov:LASE}) becomes  to $\hat{\mathfrak H}(t,\Psi(t))$
in (\ref{shapovalov:GPE-2a}). This implies  that the function
$\Phi (\vec x, t, \hbar, {\mathfrak C}(\psi))$ satisfies equation
(\ref{shapovalov:GPE-2a}) and the initial condition
(\ref{shapovalov:LASE-C}), which correlates  with
(\ref{shapovalov:CAUCHY}). Consequently, equation
(\ref{shapovalov:CAUCHY-SOL}) is  valid, and the theorem is
proved.
\end{proof}

The relationship between the  steps described above  can be shown  diagrammatically:
%%%%%%%%%%%%%%%%%%%%%%%%%%%%%%%%%%%%%%%%%%%%%%%%%%%%%%%%%%%%%%%%
%%%%%%%%%%%%%%%%%%%%%%%%%%%%%%%%%%%%%%%%%%%%%%%%%%%%%%%%%%%%%%%%%

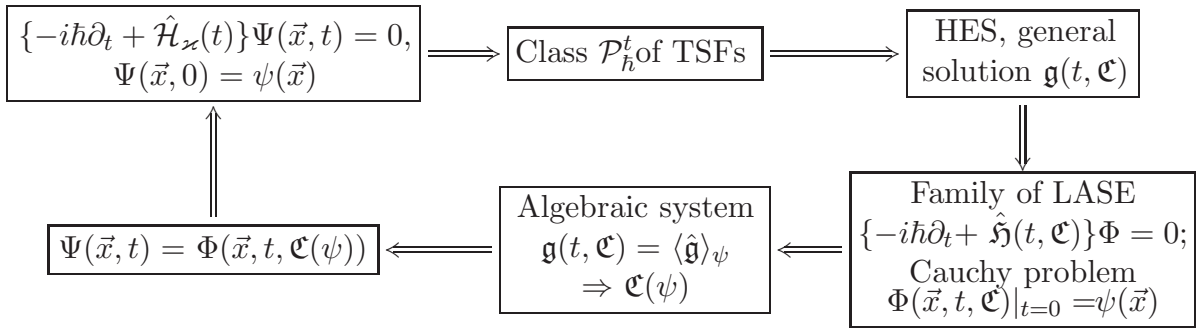
\xymatrix{
\text{\framebox[55mm][t]{\scalebox{1.1}{\parbox{48.5mm}{\centering $\{-i\hbar\partial_t+
\hat{\mathcal H}_\varkappa(t)\}\Psi(\vec x,t)=0,$\\
$\Psi(\vec x,0)=\psi(\vec x)$}}}} \ar@{=>}[r] &
\text{\framebox[34mm][t]{\scalebox{1.1}{\parbox{30mm}{Class ${\mathcal P}_\hbar^t$%\\
of TSFs }}}} \ar@{=>}[r] &
\text{\framebox[31mm][t]{\scalebox{1.1}{\parbox{27mm}{\centering HES, general\\
solution
${\mathfrak g}(t,{\mathfrak C})$
\\
}}}} \ar@{=>}[d]
%\ar[dr] \ar[dl]
\\
\text{\framebox[44mm][t]{\scalebox{1.1}{\parbox{38mm}{\centering
$\Psi(\vec x,t)=$
$\Phi(\vec x,t,{\mathfrak C}(\psi))$}
}}}\ar@{=>}[u] &
\text{\framebox[36mm][t]{\scalebox{1.1}{\parbox{31mm}{\centering
Algebraic system\\
${\mathfrak g}(t,{\mathfrak C})=$
$\langle \hat{\mathfrak g}\rangle_{\psi}$
%\
\\
$\Rightarrow$
${\mathfrak C}(\psi)$}}}} \ar@{=>}[l] &
\text{\framebox[46mm][t]{\scalebox{1.1}{\parbox{40mm}{\centering
Family of LASE\\
$\lbrace -i\hbar\partial_t +$
$\hat{\mathfrak H}(t,{\mathfrak C})\rbrace\Phi=0$;
\\ Cauchy problem\\
\!\!$\Phi(\vec x, t,{\mathfrak C})|_{t=0}=\!\!\psi(\vec x)$
}}}} \ar@{=>}[l]
}

\medskip

Theorem \ref{shapovalov:THEOR2} makes it possible to obtain a
nonlinear evolution operator for the GPE (\ref{shapovalov:GPE-2a})
in the ${\mathcal P}^t_\hbar$ class of TCFs
(\ref{shapovalov:TCF}). The evolution operator can be written as a
nonlinear integral operator using the Green function of the LASE
(\ref{shapovalov:LASE}) with constants ${\mathfrak C}$ changed by
${\mathfrak C}(\psi)$ according to  relation
(\ref{shapovalov:CAUCHY-SOL}).

\section{Nonlinear evolution operator}
\label{shapovalov:SECT-EVOLUTION}

The Green function
$G_\varkappa\big(\vec x,\vec y,t,s,{\mathfrak g}(t,{\mathfrak C}),{\mathfrak g}(s,{\mathfrak C}))$
for the Cauchy problem (\ref{shapovalov:LASE}), (\ref{shapovalov:LASE-C}) is defined by the conditions
\begin{gather}
 [-i\hbar\partial_t + \hat{\mathfrak H}(t,{\mathfrak g}(t,{\mathfrak
C}))] G_\varkappa\big(\vec x,\vec y,t,s,{\mathfrak g}
(t,{\mathfrak C}),{\mathfrak g}(s,{\mathfrak C})\big)=0,\label{shapovalov:GREEN1}\\
\lim\limits_{t\to s} G_\varkappa\big(\vec x,\vec
y,t,s,{\mathfrak g}(t,{\mathfrak C}),
{\mathfrak g}(s,{\mathfrak C})\big)=\delta (\vec x-\vec y).\label{shapovalov:GREEN2}
\end{gather}
Here the operator $\hat{\mathfrak H}(t,{\mathfrak g}(t,{\mathfrak
C}))$,  given by (\ref{shapovalov:LASE-OPR}), is quadratic in
coordinates and momenta.

 We shall seek for the required Green function
under  the simplifying assumption
\begin{gather}
\det{\mathfrak H}_{pp} (s) \neq 0.
\label{shapovalov:SIMP-ASSUM}
\end{gather}
Following, for example, \cite{shapovalov:MANKO,shapovalov:PEREL}, we obtain
\begin{gather}
G_\varkappa\big(\vec x,\vec y,t,s,{\mathfrak g}(t,{\mathfrak C}),{\mathfrak g}(s,{\mathfrak C}))
= \frac{1}{\sqrt{\det(-i2\pi\hbar\lambda_3(t,s))}} \nonumber\\
\qquad{}
\times \exp\Bigg\lbrace\frac i
\hbar\Bigg[S(t,\hbar, {\mathfrak g}(t,{\mathfrak C}))  -S(s,\hbar, {\mathfrak g}(s,{\mathfrak C}))+
\langle\vec P(t,\hbar, {\mathfrak C}),\Delta\vec
x\rangle-\langle\vec P(s,\hbar, {\mathfrak C}),\Delta\vec y \rangle\nonumber\\
\qquad-  \frac 1 2
\langle\Delta\vec y,\lambda_1(t,s)\lambda^{-1}_3(t,s) \Delta\vec y
\rangle+\langle\Delta\vec x,\lambda_3^{-1}(t,s)\Delta\vec y\rangle -\frac 1 2\langle\Delta\vec
x,\lambda_3^{-1}(t,s)\lambda_4(t, s)\Delta\vec x\rangle\Bigg] \!\Bigg\rbrace.\!\!\!\!\! \label{shapovalov:SIMP-1}
\end{gather}
Here, $\Delta \vec y= \vec y - \vec X(s,\hbar, {\mathfrak C} )$,
$n\times n$  matrices  $\lambda _k(t,s)$, $k=\overline{1,4}$, are
blocks of the block matrix ${\mathcal A}(t,s)$ of the system in
variations:
\begin{gather}
\dot{\mathcal A}=J{\mathfrak H}_{zz}(t,\hbar)A,\qquad
{\mathcal A}\Big|_{t=s}={\mathbb I}_{2n\times 2n}, \label{shapovalov:MATRICIANT}\\
 {\mathcal A}(t,s)=\begin{pmatrix} \lambda_4^\intercal(t,s)  & -\lambda_2^\intercal(t,s) \cr
-\lambda_3^\intercal(t,s) & \lambda_1^\intercal (t,s)\end{pmatrix},
\label{shapovalov:MATRICIANT-1}
\end{gather}
and
\begin{gather}
 S(t, \hbar, {\mathfrak g}(t,
{\mathfrak C}))=
 \int_0^t \Bigl\lbrace\langle
\vec P(t,\hbar, {\mathfrak C}),\dot{\vec
X}(t,\hbar, {\mathfrak C})\rangle - {\mathfrak H}(t,\hbar,{\mathfrak
C})\Bigr\rbrace dt. \label{shapovalov:EVOLUT2}
\end{gather}
Then, the following theorem is true:
\begin{theorem}
\label{shapovalov:THEOR3} Let an  operator  $\hat U_\varkappa
\big(t,s,\cdot\big)$ act on a given function $\psi(\vec x)$,
taken at an initial time~$s$, as follows:
\begin{gather}
\hat U_\varkappa\big(t,s,\psi\big)(\vec x)= \int_{{\mathbb R}^n} G_\varkappa\big(\vec
x,\vec y,t,s,{\mathfrak g}(t,{\mathfrak C}(\psi)),{\mathfrak g}(s,{\mathfrak C}(\psi))\psi(\vec y)
\,d^ny.\label{shapovalov:EVOLUT}
\end{gather}
Here
$G_\varkappa\big(\vec x,\vec y,t,s,{\mathfrak g}(t,{\mathfrak C}(\psi),
{\mathfrak g}(s,{\mathfrak C}(\psi)))\big)$
is determined by   \eqref{shapovalov:SIMP-1}, \eqref{shapovalov:EVOLUT2},
and the parameters ${\mathfrak C}(\psi)$ are obtained from
\begin{gather}
{\mathfrak g}(t,{\mathfrak C})\Big|_{t=s}={\mathfrak g}_0(\psi)=
 \frac{1}{\|\psi \|^2}\langle\psi|\hat{\mathfrak g}|
\psi\rangle.\label{shapovalov:EVOLUT3}
\end{gather}
Then, the function
\begin{gather}
 \Psi (\vec x,t)=\hat U_\varkappa\big(t,s,\psi\big)(\vec
x)\label{shapovalov:EVOLUT4}
\end{gather}
is an {\it exact solution} of the Cauchy problem for   equations
\eqref{shapovalov:GPE-2a}, \eqref{shapovalov:GPE-2} with the
initial condition  $\Psi(\vec x,t)\big|_{t=s}= \psi(\vec x)$, and
the operator  $\hat U_\varkappa\big(t,s,\cdot\big)$ is the {\sl
evolution operator} for the GPE \eqref{shapovalov:GPE-2a} that
acts on the class of trajectory concentrated functions
\eqref{shapovalov:TCF}.
\end{theorem}
For the evolution operator (\ref{shapovalov:EVOLUT}), the following properties can be
verified by direct computation.
%\bigskip
\begin{theorem}\label{shapovalov:THEOR4}
The operator
$\hat U_\varkappa^{-1}\big(t,s,\cdot\big)$,
\begin{gather}
\hat U_\varkappa^{-1}\big(t,s,\psi\big)(\vec
x)= \int_{{\mathbb R}^n}
G_\varkappa^{-1}\big(\vec x,\vec y,t,s,{\mathfrak
g}(t,(\psi)),{\mathfrak g}(s,{\mathfrak C}(\psi))\big)\psi(\vec y)
\,d^ny\nonumber \\
\phantom{\hat U_\varkappa^{-1}\big(t,s,\psi\big)(\vec x)}{} = \int_{{\mathbb R}^n}
G_\varkappa\big(\vec x,\vec y,s,t,{\mathfrak g}(s,{\mathfrak C}
(\psi)),{\mathfrak g}(t,(\psi))\big)\psi(\vec y) \,d^ny,
\label{shapovalov:EVOLUT5}
\end{gather}
is the left inverse operator to $\hat U_\varkappa
\big(t,s,\cdot\big)$ of \eqref{shapovalov:EVOLUT}, so that
\begin{gather}
\hat U_\varkappa^{-1}\big(t,s,\hat
U_\varkappa\big(t,s,\psi\big)\big)(\vec x)=\psi(\vec x),\qquad \psi\in{\mathcal P}_\hbar^0.
\label{shapovalov:EVOLUT-INV}
\end{gather}
\end{theorem}
\begin{corollary}
\label{shapovalov:COROL1}  For a partial  solution  $\Psi(\vec
x,t)$ of the GPE \eqref{shapovalov:GPE-2a}, we have
\begin{gather}
\hat U_\varkappa \big(t,s,\hat
U_\varkappa ^{-1}\big(t,s,\Psi(t)\big)\big)(\vec x)=\Psi(\vec x,t).
\label{shapovalov:EVOLUT-INV1}
\end{gather}
\end{corollary}

\begin{proof}
Indeed, according to  Theorem \ref{shapovalov:THEOR3} we have
$\Psi(\vec x,t)=\hat U_\varkappa \big(t,s,\psi\big)(\vec x)$,
$\psi(\vec x)=\Psi(\vec x,t)\big|_{t=s}$. In view of
(\ref{shapovalov:EVOLUT-INV}), the left-hand side of equation
(\ref{shapovalov:EVOLUT-INV1}) can be written as
\begin{gather*}
\hat U_\varkappa\big(t,s,\hat
U_\varkappa^{-1}\big(t,s,\Psi(t)\big)\big)(\vec x)=
\hat U_\varkappa\big[t,s,\hat U_\varkappa^{-1}\big(t,s,\hat
U_\varkappa\big(t,s,\psi\big)\big)\big](\vec x)
=\hat U_\varkappa\big(t,s,\psi\big)(\vec x)=\Psi(\vec x,t).
\end{gather*}
Thus, the statement is proven.
\end{proof}

\begin{theorem}\label{shapovalov:THEOR5} The operators
$ \hat U_\varkappa\big(t,\cdot\big)= \hat
U_\varkappa \big(t,0,\cdot\big)$ possess the group
property
\begin{gather}
\hat U_\varkappa\big(t+s,\psi\big)(x)=\hat
U_\varkappa\big(t,\Psi(s)\big)(\vec x), \qquad
\Psi(\vec x,s)=\hat U_\varkappa\big(s,\psi\big)(\vec
x).\label{shapovalov:GROUP}
\end{gather}
\end{theorem}

\begin{proof} 
The functions $\Psi(\vec x,t)=\hat
U_\varkappa\big(t+s,\psi\big)(\vec x)$ and $\Psi(\vec x,t)$ of
(\ref{shapovalov:EVOLUT4}) are partial solutions of equation
(\ref{shapovalov:LASE}) with the same trajectory ${\mathfrak
g}(t,{\mathfrak C})$ in the extended phase space, and
(\ref{shapovalov:GROUP}) is valid for the evolution operator of
the linear equation (\ref{shapovalov:LASE}). Then, it is also valid
for ${\mathfrak g}(t,{\mathfrak C}(\psi))$ corresponding to the
nonlinear evolution operator (\ref{shapovalov:EVOLUT}).

Substituting $t+s\to t$ in (\ref{shapovalov:GROUP}), we find
\begin{gather}
\hat U_\varkappa\big(t,\psi\big)(\vec x)=\hat
U_\varkappa\big(t-s,\Psi(s)\big)(\vec x), \label{shapovalov:GROUP1}\\
\Psi(\vec x,s)\Big|_{s=0}=\psi(\vec x)\label{shapovalov:GROUP2}
\end{gather}
for a solution $\Psi(\vec x,t)$ of the Cauchy problem for the GPE
(\ref{shapovalov:GPE-2a}) with the initial condition (\ref{shapovalov:GROUP2}).
\end{proof}

\section{Symmetry and nonlinear superposition}

\subsection*{Symmetry operators}

A symmetry operator, by definition, maps a solution of an equation
into another solution of this equation.
Direct finding of symmetry operators for a given nonlinear equation is an
intricate problem because of the nonlinearity of the  determining equations.

It is rare for this problem to be solved  (see, for example,
\cite{shapovalov:PUKHNACHEV}). The  symmetry analysis of
differential equations deals mainly with generators of
one-parametric families  of symmetry operators (symmetries of an
equation) determined by linear equations
\cite{shapovalov:OVS,shapovalov:IBRAGIM,shapovalov:FUSCH-SS,shapovalov:FUSCH-N,shapovalov:OLVER}.
Using the evolution operator $\hat U_\varkappa
(t,s,\cdot)$ given by (\ref{shapovalov:EVOLUT}), we can
formulate a general form  for symmetry operators of
the Gross--Pitaevskii equation (\ref{shapovalov:GPE-2a}).

Let $\hat {\textsf{a}}$ be an operator acting in
 ${\mathcal P}^0_\hbar$, ($\hat {\textsf{a}}:{\mathcal P}^0_\hbar
\to{\mathcal P}^0_\hbar$) and $\Psi(\vec x,t)$ is an arbitrary function
of the ${\mathcal P}^t_\hbar$ class  ($\Psi(\vec x,t)\in{\mathcal P}^t_\hbar$).
Consider an operator $\hat {\textsf{A}}(\cdot)$, such that
\begin{gather}
\Phi(\vec x,t)=\hat {\textsf{A}}\big(\Psi(t)\big)(\vec x)=\hat
U_\varkappa\big(t,\hat {\textsf{a}}\, \hat
U_\varkappa^{-1}\big(t,\Psi(t)\big)\big)(\vec x). \label{shapovalov:SYMM}
\end{gather}
If  $\Psi(\vec x,t)$ is a solution of the GPE
(\ref{shapovalov:GPE-2a}), then  $\Phi(\vec x,t)$ is also a
solution of equation (\ref{shapovalov:GPE-2a}). This follows
immediately from Theorems \ref{shapovalov:THEOR3} and
\ref{shapovalov:THEOR4}, and Corollary \ref{shapovalov:COROL1}.

Thus, the operator  {\rm $\hat {\textsf{A}}(\cdot)$}
determined by (\ref{shapovalov:SYMM}) is a symmetry operator for
the GPE (\ref{shapovalov:GPE-2a}).

Assume  now that  operator $\hat {\textsf{b}}$ and its operator
exponent $\exp(\alpha\hat {\mathfrak b})$ act in the ${\mathcal
P}^0_\hbar$ class, i.e., $\hat {\textsf{b}}:{\mathcal P}^0_\hbar
\to{\mathcal P}^0_\hbar$ and $\exp(\alpha\hat
{\textsf{b}}):{\mathcal P}^0_\hbar \to{\mathcal P}^0_\hbar$, where
$\alpha$ is a parameter.

Define a one-parametric family of operators
$\hat {\textsf{B}}(\alpha,\cdot)$
via their action on an arbitrary function
$\Psi(\vec x,t)\in{\mathcal P}^t_\hbar$  as
\begin{gather}
\hat {\textsf{B}}\big(\alpha,\Psi(t)\big)(\vec x)=\hat
U_\varkappa\big(t,\exp\lbrace\alpha\hat {\textsf{b}}\rbrace \hat
U_\varkappa^{-1}\big(t,\Psi(t)\big)\big)(\vec x). \label{shapovalov:SYMM1}
\end{gather}
By analogy with the aforesaid, the operators
$\hat {\textsf{B}}(\alpha,\cdot)$
constitute a one-parametric family of the symmetry operators
of equation (\ref{shapovalov:GPE-2a}).

It is easy to verify the group property
\begin{gather}
\hat {\textsf{B}}\big(\alpha+\beta,\Psi(t)\big)(\vec x)=\hat
{\textsf{B}}\Big(\alpha,\hat
{\textsf{B}}\big(\beta,\Psi(t)\big)\Big)(\vec x),
\qquad \forall\;\Psi(\vec x,t)\in{\mathcal P}^t_\hbar.
\label{shapovalov:SYMM2}
\end{gather}
Differentiating
(\ref{shapovalov:SYMM1})
with respect to
the parameter $\alpha$, we obtain for $\alpha=0$
\begin{gather}
\hat{\textsf{C}}\big(\Psi(t)\big)(\vec x)= \frac{d}{d \alpha}\hat
{\textsf{B}}\big(\alpha,\Psi(t)\big)(\vec x)\Big|_{\alpha=0}= \frac{d}{d
\alpha}\hat U_\varkappa\big(t,\exp\lbrace\alpha\hat {\textsf{b}}\rbrace \hat
U_\varkappa^{-1}\big(t,\Psi(t)\big)\big)(\vec x)\Big|_{\alpha=0}.
\label{shapovalov:SYMM3}
\end{gather}
The operator $\hat{\textsf{C}}(\cdot)$ determined by
(\ref{shapovalov:SYMM3}) is a generator of the one-parametric family of symmetry operators
(\ref{shapovalov:SYMM1}).

Note that the operator $\hat{\textsf{C}}(\cdot)$ is not a symmetry operator
for  equation (\ref{shapovalov:GPE-2a}) since the parame\-ters~${\mathfrak C}$ in the
evolution operator $\hat U_\varkappa(t,\cdot)$
(\ref{shapovalov:EVOLUT}) depend on $\alpha$.
Indeed, the parameters  $(\mathfrak C)$ found  from equation
(\ref{shapovalov:EVOLUT3})
\begin{gather*}
{\mathfrak g}(t,{\mathfrak C})\Big|_{t=0}=\langle\exp\lbrace\alpha\hat {\textsf{b}}\rbrace\phi|
\hat{\mathfrak g}| \exp\lbrace\alpha\hat {\textsf{b}}\rbrace\phi\rangle, \qquad
\phi(\vec x)=\hat U_\varkappa^{-1}\big(t,\Psi(t)\big)(\vec x),\qquad
\Psi(\vec x,t)\in{\mathcal P}^t_\hbar
\end{gather*}
include the parameter $\alpha$ explicitly.
Therefore,
equation  (\ref{shapovalov:SYMM3}) includes the derivatives of
the evolution operator $ \hat U_\varkappa(t,\cdot)$
with respect to the parameters ${\mathfrak C}$, and
(\ref{shapovalov:SYMM3}) is different in form from the  symmetry operator
(\ref{shapovalov:SYMM}).

\subsection*{Nonlinear superposition}

The nonlinear superposition principle for the GPE
(\ref{shapovalov:GPE-2a}) can be formulated
 in terms of the evolution operator

Let
\begin{gather}
\Psi_1(\vec x,t)=\hat U_\varkappa\big(t,\psi_1\big)(\vec
x),\qquad
\Psi_2(\vec x,t)=\hat U_\varkappa\big(t,\psi_2\big)(\vec
x)\, \in {\mathcal P}^t_\hbar
\label{shapovalov:SUPERPOS}
\end{gather}
be two partial solutions to the GPE (\ref{shapovalov:GPE-2a})
corresponding to the initial functions
$\psi_1(\vec x),\psi_2(\vec x)$ $\in {\mathcal P}^0_\hbar$, respectively.

Then, the function
\[
\Psi(\vec x,t)=\hat U_\varkappa\big(t,c_1\psi_1+ c_2\psi_2\big)(\vec
x)
\]
is a solution of  equation (\ref{shapovalov:GPE-2a}) which corresponds to the initial function
$c_1\psi_1(\vec x)+ c_2\psi_2(\vec x)$, \mbox{$c_1, c_2\in {\mathbb R}^1$}.
 Therefore,
\begin{gather}
\Psi(\vec x,t)=\hat U_\varkappa\big(t,
c_1\hat U_\varkappa^{-1}\big(t,\Psi_1(t)\big)+
c_2\hat U_\varkappa^{-1}\big(t,\Psi_2(t)\big)\big)(\vec x)
\label{shapovalov:SUPERPOS1}
\end{gather}
is a solution of equation (\ref{shapovalov:GPE-2a}) which corresponds to the solutions
$\Psi_1(\vec x,t)$, $\Psi_2(\vec x,t)$ of the form~(\ref{shapovalov:SUPERPOS}), and
equation (\ref{shapovalov:SUPERPOS1}) is  the {\it superposition principle}
for the GPE (\ref{shapovalov:GPE-2a}).

\section{Examples}

\subsection*{3D case}

Consider equations (\ref{shapovalov:GPE}) and (\ref{shapovalov:VTE-2})
in a 3-dimensional space with operators $\hat {\mathcal H}(t)$,
$\hat V(t,\Psi )$ of the form
\begin{gather}
\hat {\mathcal H}(t)= \frac{1}{2m}\Big( \hat{\vec p}-
 \frac{e}{c}\vec A(\vec x,t)\Big)^2-e \langle \vec E(t),\vec x\rangle
+ \frac{k}{2}\vec x^2, \label{shapovalov:EXAMP1-1}\\
 V(\hat z,\hat w,t) =V(\vec x-\vec y)=V_0\Big(1- \frac{1}{2\gamma^2}(\vec x-\vec y)^2\Big).
\label{shapovalov:EXAMP1-2}
\end{gather}
%%%%%%%%%%%%%%%%%
The external field  in the linear operator
(\ref{shapovalov:EXAMP1-1}) is the superposition of a constant
magnetic field $\vec H=(0,0,H)$ with vector potential $\vec
A= \frac{1}{2}\vec H\times \vec x$, an electric field
$\vec E(t)=(E\cos\omega t,E\sin\omega t,0)$ periodic in time with
frequency $\omega$, and the field of an isotropic oscillator with
potential $ \frac{k}{2}\vec x ^2$, $k>0$. The operator
$\hat {\mathcal H}(t)$ is the same  as in
\cite{shapovalov:KIEV03}, and $V(\vec x-\vec y)$ is  obtained from
the Taylor expansion of the Gaussian potential $V(\vec x-\vec y)=$
$V_0\exp\left [- \frac{(\vec x-\vec y)^2}{2\gamma^2}
\right ].$ In notations (\ref{shapovalov:GPE-2a})--(\ref{shapovalov:GPE-2b}), we have
\begin{gather}
{\mathcal H}_{zz}=\left(
\begin{array}{ll}{\mathcal H}_{pp} & {\mathcal H}_{px}\\
{\mathcal H}_{xp} & {\mathcal H}_{xx}
\end{array}\right),\qquad
{\mathcal H}_{pp}= \frac{1}{m}{\mathbb I}_{3\times 3},
\qquad {\mathcal H}_{xx}=\mbox{diag}\big\lbrace m\omega_1^2,
m\omega_1^2,
m\omega_2^2 \big\rbrace ; \nonumber \\
 {\mathcal H}_{z}=(\vec{{\mathcal H}}_{p},\vec{{\mathcal
H}}_{x})^\intercal, \quad   \vec{\mathcal H}_{p}=0,\qquad
\vec{\mathcal H}_{x}=-e \vec E(t);\label{shapovalov:PARAM}  
\end{gather}
the nonzero elements of the $3\times 3$  matrix ${\mathcal
H}_{px}(={\mathcal H}_{xp}^\intercal)$ are: ${\mathcal
H}_{p_1x_2}=-{\mathcal
H}_{p_2x_1}= \frac{1}{2}\omega_H$. Here,
$\omega_H= \frac{eH}{mc}$ is a {\it cyclotron
frequency}, $\omega_{nl}(v)^2=$ $ \frac{|\tilde\varkappa v|}{m}$
is a ``nonlinear frequency'',
$\omega_0^2= \frac{k}{m}$,
$\omega_1^2=\omega_0^2+\big( \frac{\omega_H}{2}\big)^2
-\zeta(V_0)\omega_{nl}(\eta)^2$,
$\omega_2^2=\omega_0^2 -\zeta(V_0)\omega_{nl}(\eta)^2$,
$\zeta(a)={\mbox{sign}}(\tilde\varkappa a)$, $a\in {\mathbb R}^1$,
$\eta= \frac{V_0}{\gamma^2}$. The nonzero
matrices in (\ref{shapovalov:QUAD-2}) are:
$W_{xx}=W_{yy}=-W_{xy}=- \frac{1}{\gamma^2}V_0{\mathbb
I}_{3\times 3}$.

For the block matrix  ${\mathcal A}(t)$ of the form
(\ref{shapovalov:MATRICIANT-1}), the matrices $\lambda_1,
\lambda_2, \lambda_3, \lambda_4$ are of order $3\times 3$ and they
can be written as
\begin{gather*}
\lambda_1=\lambda_4= \frac{d \hat{\mathbf r}(t)}{dt}\hat{\mathbf u}(t),
\qquad
\lambda_2=-m \frac{d^2 \hat{\mathbf r}(t)}{dt^2}\hat{\mathbf u}(t),
\qquad
\lambda_3=- \frac{1}{m}\hat{\mathbf r}(t)\hat{\mathbf u}(t),
\end{gather*}
where
\begin{gather*}
\hat{\mathbf r}(t)={\mbox{diag}}\left\lbrace
 \frac{\sin(\omega_1t)}{\omega_1},
 \frac{\sin(\omega_1t)}{\omega_1},
 \frac{\sin(\omega_2t)}{\omega_2} \right\rbrace,\qquad
\hat{\mathbf u}(t)= \left(
\begin{array}{lll} \cos( \frac{1}{2}\omega_Ht) &
-\sin( \frac{1}{2}\omega_Ht)&0\\
\sin( \frac{1}{2}\omega_Ht) &
\cos( \frac{1}{2}\omega_Ht)&0\\
0&0& 1
\end{array}\right).\!
\end{gather*}
Here $\hat{\mathbf u}(t)$ is an  $SO(3)$  matrix, $\hat{\mathbf
u}^\intercal(t)\hat{\mathbf u}(t)=$ $\hat{\mathbf
u}(t)\hat{\mathbf u}^\intercal(t)= {\mathbb I}_{3\times 3}$.
Denote by $(\vec P(t,\hbar, {\mathfrak C})$, $\vec
X(t,\hbar,{\mathfrak C}))$ the general solution of the system
(\ref{shapovalov:HES-ABST}) with  (\ref{shapovalov:PARAM}), for
which we use for short the notation $(\vec P(t), \vec X(t))$.
 We use similar notation for (\ref{shapovalov:EVOLUT2}), $S(t, \hbar,
{\mathfrak g}(t,{\mathfrak C}))=$ $S(t)$.

Let us introduce the functions
\begin{gather}
{\mathcal G}(\Delta x,\Delta y, t, s, P(t), P(s), \omega, \omega_H, {\mathfrak C})=
\sqrt{ \frac{m\omega}{2i\pi \hbar\sin(\omega(t-s))}}
\exp\left\lbrace \frac{i}{\hbar}(P(t)\Delta x-P(s)\Delta y)
\right\rbrace \nonumber \\
\quad {}\times\exp\left\lbrace \frac{i\omega m}{2\hbar\sin(\omega(t-s))}
\left(\cos(\omega(t-s))(\Delta x^2+\Delta y^2)-
2\cos( \frac{\omega_H}{2}(t-s))\Delta x\Delta y
\right)
\right\rbrace
\label{shapovalov:GREEN-1}
\end{gather}
and
\begin{gather}
{\mathcal G}(\Delta x,\Delta y, t, s, P(t), P(s),
\omega,{\mathfrak C})= {\mathcal G}(\Delta x,\Delta y, t, s, P(t),
P(s), \omega, \omega_H, {\mathfrak C})\vert_{\omega_H=0}.
\label{shapovalov:GREEN-2}
\end{gather}
Then, the Green function (\ref{shapovalov:SIMP-1}) reads
\begin{gather}
G_\varkappa\big(\vec x,\vec y,t,s,{\mathfrak g}(t,{\mathfrak C}),{\mathfrak g}(s,{\mathfrak C}))=
{\mathcal G}(\Delta x_1,\Delta y_1, t, s, P_1(t), P_1(s), \omega_1,\omega_H, {\mathfrak C})
\nonumber\\
\quad{}\times{\mathcal G}(\Delta x_2,\Delta y_2, t, s, P_2(t), P_2(s),
\omega_1,\omega_H, {\mathfrak C}) \, {\mathcal G}(\Delta
x_3,\Delta y_3, t, s, P_3(t), P_3(s), \omega_2,{\mathfrak C})
 \nonumber\\
\quad{}\times
\exp\left\lbrace \frac{i}{\hbar}[S(t)-S(s)]\right\rbrace
\exp\left\lbrace \frac{i}{\hbar}
\left(-m\omega_1 \frac{\sin( \frac{\omega_H}{2}(t-s))}
{\sin(\omega_1(t-s))}(\Delta x_1\Delta y_2-\Delta x_2\Delta y_1)\right)
\right\rbrace.\label{shapovalov:GREEN-LIN} 
\end{gather}

The nonlinear evolution operator (\ref{shapovalov:EVOLUT}) is
constructed with the use of (\ref{shapovalov:GREEN-LIN}) in which
${\mathfrak C}$ are determined by (\ref{shapovalov:EVOLUT3}).

\subsection*{1D case}

To demonstrate symmetry operators in explicit form
consider the  one-dimensional case of the GPE (\ref{shapovalov:GPE-2a})
following \cite{shapovalov:LTS}.

The operators (\ref{shapovalov:EXAMP1-1}) and  (\ref{shapovalov:EXAMP1-2})
 for $n=1$ take the form
\begin{gather}
 {\mathcal H}(\hat z,t)= \frac{\hat p^2}{2m} +
\frac{kx^2}{2}-eEx\cos\omega t, \quad
\quad 
V(\hat z, \hat w, t)=\frac 12
\big[ax^2+2bxy+cy^2\big]. \label{shapovalov:EXAMP2-2}
\end{gather}
Here, $\vec x=x\in {\mathbb R}^1$, $\vec y=y\in {\mathbb R}^1$;
$k>0$, $m$, $e$, $E$, $a$, $b$, and  $c$  are
parameters of the potential.

In notations (\ref{shapovalov:GPE-2a})--(\ref{shapovalov:GPE-2b}), we have
$ {\mathfrak H}(t,\hbar,{\mathfrak g}(t,{\mathfrak
C}))= $ $ \frac{1}{2m}{P^2(t,{\mathfrak C})}+$
$ \frac{1}{2}kX^2(t,{\mathfrak
C})- $ $eEX(t,\mathfrak C)\cos \omega t +$ $
 \frac{\tilde\varkappa}2c\sigma_{xx}(t,{\mathfrak
C})+ $ $ \frac{\tilde\varkappa}2(a+2b+c)X^2(t,{\mathfrak C}) $,
\begin{gather*}
{\mathfrak H}_z(t,{\mathfrak g}(t,{\mathfrak C}))= \left(
\begin{array}{c}
 \frac{1}{m}P(t,{\mathfrak C})\\%[6 pt]
  m\tilde\Omega^2X(t,{\mathfrak C})-eE\cos \omega t\end{array}\right),%\cr &&
\qquad{\mathfrak H}_{zz}(t,{\mathfrak g}(t,{\mathfrak C})))=\left(
\begin{array}{cc}
 \frac{1}{m}&0 \\
0 & m\Omega^2
\end{array}\right).\nonumber
\end{gather*}
The Green function of the linear associated
Schr\"odinger equation (\ref{shapovalov:LASE})
follows directly from (\ref{shapovalov:GREEN-1}) and
(\ref{shapovalov:GREEN-LIN}). It reads
\begin{gather}
G_\varkappa\big( x, y,t,s,{\mathfrak g}(t,{\mathfrak C}),{\mathfrak g}(s,{\mathfrak C}))
={\mathcal G}(\Delta x,\Delta y, t, s, P(t), P(s), \Omega, {\mathfrak C})
\exp\left\lbrace \frac{i}{\hbar}[S(t)-S(s)]\right\rbrace .
\label{shapovalov:GREEN-LIN-1D}
\end{gather}
Here, $\omega_0=\sqrt{k/m}$ and $\Omega^2=$
$\omega_0^2+\zeta(a)\omega_{nl}^2(a)$. The evolution operator
$\hat U_{\varkappa}(t,s, \cdot)$ (\ref{shapovalov:EVOLUT}) is
determined by (\ref{shapovalov:GREEN-LIN-1D}), where the constants
${\mathfrak C}$ are changed by ${\mathfrak C}(\psi)$ determined by
equation (\ref{shapovalov:CONS-2}): ${\mathfrak g}(s,{\mathfrak C})=$
${\mathfrak g}({\mathfrak C})= \frac{1}{\|\psi\|^2}\langle\psi|\hat{\mathfrak
g}|\psi\rangle$, ${\mathfrak g}({\mathfrak C})=(P({\mathfrak
C}),X({\mathfrak C})$, $\sigma_{pp}({\mathfrak C}),
\sigma_{px}({\mathfrak C})$, $\sigma_{xx}({\mathfrak C}))$. The
LASE (\ref{shapovalov:LASE}), (\ref{shapovalov:LASE-OPR}) is known
to have a partial solution in  Gaussian form. For the operators
(\ref{shapovalov:EXAMP2-2}), this solution can be obtained as
\begin{gather*}
\Phi_0^{(0)}(x,t,{\mathfrak g}(t,{\mathfrak C}))\!=\!
\sqrt[4]{ \!\frac{m\Omega}{\pi\hbar}}e^{-i\Omega t/2}
\exp\biggl\{ \frac{i}{\hbar}\Big(\!
S(t,\hbar,{\mathfrak g}(t,{\mathfrak C}))\!+\! P(t,{\mathfrak
C})\Delta x\!\Big)\!-\! \frac{m\Omega}{2\hbar}\Delta x^2\biggr\}.
\end{gather*}
Setting $\psi(x)=$
$\Phi_0^{(0)}(x,0,{\mathfrak g}(0,{\mathfrak C}))\equiv $ $\Phi_0^{(0)}(x,0) $
in (\ref{shapovalov:CONS-2}), we can find the parameters
${\mathfrak C}$ in the form
${\mathfrak C}(\psi)={\mathfrak C}_0=$ ${\mathfrak C}(\Phi_0^{(0)}(0))=$
$\bigg( \frac{p_0}{m\tilde\Omega},x_0,0,0,
 \frac{\hbar}{2m\Omega}\bigg)^\intercal$,
where $\tilde\Omega^2=$ $\omega_0^2+\zeta(a+b)\omega_{nl}^2(a+b)$,
$p_0=\langle \Phi_0^{(0)}|\hat p |\Phi_0^{(0)} \rangle/\|\Phi_0^{(0)}\|^2$, and
$x_0=\langle \Phi_0^{(0)}| x |\Phi_0^{(0)} \rangle/\|\Phi_0^{(0)}\|^2$.
Then,
\begin{gather}
\Psi_0^{(0)}(x,t,{\mathfrak C}_0)= \Phi_0^{(0)}(x,t,{\mathfrak g}(t,{\mathfrak C(\psi)}))
\end{gather}
is a partial solution of the 1-dimensional GPE (\ref{shapovalov:GPE-2a}).

 Let us change the operators $\hat {\textsf{a}}$ in   relation
(\ref{shapovalov:SYMM}) by
the operators $\hat a^+$ and $\hat a$ of the form
\begin{gather*}
 \hat a=\frac1{\sqrt{2\hbar m\Omega}}\big[
\Delta\hat p_0-im\Omega\Delta x_0\big], \qquad \hat
a^+=\frac1{\sqrt{2\hbar m\Omega}}\big[ \Delta\hat
p_0+im\Omega\Delta x_0\big],
\end{gather*}
where $\Delta\hat p_0=-i\hbar\partial_x-p_0$ and $\Delta
x_0=x-x_0$. Then, the operators $\widehat A^{(\pm)}(\cdot)$
determined by the relations
\begin{gather*}
\widehat A^{(+)}\big(\Psi(t)\big)(x)= \widehat
U_\varkappa\big(t,\hat a^+\,
\widehat U_\varkappa^{-1}\big(t,\Psi(t)\big)\big)(x),\qquad\!\!
\widehat A^{(-)}\big(\Psi(t)\big)(x)= \widehat
U_\varkappa\big(t,\hat a\, \widehat
U_\varkappa^{-1}\big(t,\Psi(t)\big)\big)(x),\!
\end{gather*}
where $\Psi(x,t)\in{\mathcal P}^t_\hbar$, are the symmetry
operators of equation (\ref{shapovalov:GPE-2a}). With the
operators $\widehat A^{(\pm)}(\cdot)$, we obtain the set of
solutions for the GPE
\begin{gather}
\Psi_n^{(0)}(x,t,{\mathfrak g}(t,{\mathfrak C}_n))=
\frac{1}{\sqrt{n!}} \bigl(\widehat A^{(+)}(\cdot)\bigr)^n
\Psi_0^{(0)}(x,t,{\mathfrak g}(t,{\mathfrak
C}_0)) \nonumber\\
\phantom{\Psi_n^{(0)}(x,t,{\mathfrak g}(t,{\mathfrak C}_n))}{}= \frac{i^n}{\sqrt{n!}}
e^{-in\Omega t} \biggl( \frac{1}{\sqrt{2}}\biggr)^n
H_n\biggl(\sqrt{  \frac{m\Omega}{\hbar}}\Delta
x\biggr) \Psi_0^{(0)}
(x,t,{\mathfrak g}(t,{\mathfrak C}_n)),
\label{shapovalov:EXAMP2-3}
\end{gather}
where $n\in {\mathbb Z}^+$, $H_n(\xi)$ are Hermite polynomials and
\begin{gather*}
 {\mathfrak C^T_n}(\psi)\!=\!\bigg(0, \frac{e
E}{m(\tilde\Omega^2-\omega^2)}, 0, 0, \frac{\hbar(2n+1)}{2m\Omega}
\bigg)^\intercal.
\end{gather*}
It can be verified that
\begin{gather*}
\widehat A^{(+)}\big(\Psi_n^{(0)}(t,{\mathfrak g}
(t,{\mathfrak C}_n))\big)(x)
=\sqrt{n+1}\,\Psi_{n+1}^{(0)}(x,t,{\mathfrak g}(t,{\mathfrak C}_{n+1})),\\
\widehat A^{(-)}\big(\Psi_n^{(0)}(t,{\mathfrak g}
(t,{\mathfrak C}_n))\big)(x)
=\sqrt{n}\,\Psi_{n-1}^{(0)}(x,t,{\mathfrak g}(t,{\mathfrak
C}_{n-1}))
\end{gather*}
and the operators
 $\widehat A^{(\pm)}(\cdot)$ are nonlinear analogs
 of  ``creation--annihilation''  operators.
It can be noted that the functions $\Psi_n^{(0)}(x,t,{\mathfrak g}(t,{\mathfrak C}_n))$ satisfy the
quasi-periodic condition
\begin{gather*}
\Psi_n^{(0)}(x,t+T,{\mathfrak C}^T_n)=e^{-i{\mathcal E}_n T/\hbar}
\Psi_n^{(0)}(x,t,{\mathfrak C}^T_n),
\end{gather*}
where ${\mathcal E_n}$ is the quasi-energy. The spectrum of quasi-energies is given
by the relation
\begin{gather*}
{\mathcal E}_n= 
-\frac{e^2E^2}{2m(\tilde\Omega^2-\omega^2)}
-\frac{e^2E^2[\omega^2-\omega^2_0-\zeta(a+2b+c)\omega_{\rm nl}^2(a+2b+c)]}
{4m(\tilde\Omega^2-\omega^2)^2}\\
\phantom{{\mathcal E}_n=}{}+\hbar\Big(\Omega+ \frac{\tilde\varkappa c}{2m\Omega}\Big)
\Big(n+\frac12\Big).
\end{gather*}

\section{Concluding remarks}

The evolution operator (\ref{shapovalov:EVOLUT}) obtained in
Section \ref{shapovalov:SECT-EVOLUTION} in explicit form enables
one to look in a~new fashion at the problem of construction of
semiclassically concentrated solutions to the nonlocal Gross--Pitaevskii 
equation (\ref{shapovalov:GPE}). In particular, the semiclassical
asymptotics constructed in~\cite{shapovalov:BTS1,shapovalov:BTS2}
can be presented in more compact  and visual form.
In addition, the evolution operator leads to the nonlinear superposition
principle (\ref{shapovalov:SUPERPOS1}) and to the general form of symmetry
operators~(\ref{shapovalov:SYMM}). The latter can be used for study of symmetries of
the Gross--Pitaevskii equation under consideration as long as direct finding of the
symmetries
by means of solution of the determining equations is a nontrivial problem.

\subsection*{Acknowledgements}
The work was supported by  the President of the
Russian Federation, Grant  No NSh-1743.2003.2.

 \LastPageEnding

\begin{thebibliography}{999}

\footnotesize

\bibitem{shapovalov:CORNELL} Cornell~E.A., Wieman~C.E.,
Nobel lecture: Bose--Einstein condensation in a dilute gas, the
first 70 years some recent experiments, {\it Rev. Mod. Phys.},
2002, V.74,  875--893. \\
Ketterle W., Nobel lecture: When atoms
behave as waves: Bose--Einstein condensation and the atom laser,
{\it Rev. Mod. Phys.}, 2002, V.74,  1131--1151.

\bibitem{shapovalov:PITAEVSKII-1} Pitaevskii~L.P., Vortex lines in
an imperfect Bose gas, {\it Zh. Eksper. Teor. Fiz.},
1961, V.40, 646--651 (in Russian).

\bibitem{shapovalov:GROSS} Gross E.P., Structure of a quantized vortex
in boson systems, {\it Nuovo Cimento}, 1961, V.20, N~3,
454--477.


\bibitem{shapovalov:KIVSHAR}  Kivshar~Y.S.,   Pelinovsky~D.E.,
 Self-focusing and transverse instabilities of solitary waves,
{\it Phys. Rep.}, 2000, V.331, N~4, 117--195.

\bibitem{shapovalov:BANG} Bang~O., Krolikowski~W.,  Wyller~J., Rasmussen~J.J.,
Collapse arrest and soliton stabilization in nonlocal nonlinear media,
nlin.PS/0201036.


\bibitem{shapovalov:ZAKHAROV-1} Zakharov~V.E.,  Shabat~A.B.,
Exact theory of two-dimensional self-focusing and one-dimensional
self-modulation of waves in nonlinear media, {\it Zh. Eksper. Teor. Fiz.},
 1971, V.61, 118--134 (English transl.: {\it Sov. Phys. JETP}, 1971, V.34, 62--69).

\bibitem{shapovalov:ZAKHAROV-2}  Zakharov~V.E.,  Manakov~S.V.,
Novikov~S.P., Pitaevsky~L.P.,
Theory of solitons: The inverse scattering method,
Moscow,  Nauka, 1980 (English transl.: New York, Plenum,
1984).

\bibitem{shapovalov:OVS} Ovsjannikov~L.V., Group analysis of differential
equations, Moscow, Nauka,  1978 (English transl.: New York, Academic Press,   1982).

\bibitem{shapovalov:IBRAGIM}  Anderson~R. L.,  Ibragimov~N.H.,
 Lie--B\"acklund transformations in applications,  Philadelphia, SIAM, 1979.

\bibitem{shapovalov:OLVER} Olver~P.J., Application of Lie groups to
differential equations,  New York, Springer,  1986.

\bibitem{shapovalov:FUSCH-SS}
 Fushchich~W.I.,  Shtelen~W.M.,  Serov~N.I., Symmetry analysis and
exact solutions of equations of nonlinear mathematical physics,
 Dordrecht, Kluwer, 1993.

\bibitem{shapovalov:FUSCH-N} Fushchich~W.I., Nikitin ~A.G.,
Symmetries of equations of quantum mechanics, New York, Allerton Press Inc.,
 1994.


\bibitem{shapovalov:BTS1}  Belov~V.V.,  Trifonov~A.Yu.,
 Shapovalov~A.V., The trajectory-coherent approximation and
the system of moments for the Hartree type equation,
{\it Int.~J. Math. and Math. Sci.}, 2002, V.32, N~6, 325--370.

\bibitem{shapovalov:BTS2}  Belov~V.V.,  Trifonov~A.Yu., Shapovalov~A.V.,
 Semiclassical trajectory-coherent approximation for the
Hartree type equation, {\it Teor. Mat. Fiz.}, 2002, V.130, N~3,
460--492 (English transl.: {\it  Theor. Math. Phys.},
2002, V.130, N~3, 391--418).

\bibitem{shapovalov:KIEV03} Shapovalov~A.V., Trifonov~A.Yu., Lisok~A.L.,
Semiclassical approach to the geometric phase theory for
the Hartree type equation,
in Proceedinds of Fifth International Conference ``Symmetry in Nonlinear Mathematical Physics"
(June 23--29, 2003, Kyiv),
Editors A.G.~Nikitin, V.M.~Boyko, R.O. Popovych and I.A. Yehorchenko, {\it Proceedings of Institute of Mathematics},
Kyiv, 2004, V.50, Part~3, 1454--1465.


\bibitem{shapovalov:LTS} Lisok~A.L., Trifonov~A.Yu., Shapovalov~A.V.,
The evolution operator of the Hartree-type equation with 
a~quadratic potential, {\it J. Phys. A: Math. Gen.}, 2004, V.37,
4535--4556.

\bibitem{shapovalov:KARASEVMASLOV} Karasev~M.V.,  Maslov~V.P.,
Nonlinear Poisson brackets: geometry and quantization, Moscow, Nauka,
 1991 (English transl.:  Nonlinear Poisson
brackets: geometry and quantization, Ser. Translations of
Mathematical Monographs, V.119,  Providence, RI, Amer. Math.  Soc.,
 1993).

\bibitem{shapovalov:MASLOV-1}  Maslov~V.P., The complex WKB method for
nonlinear equations, Moscow, Nauka,  1977 (English transl.:
The complex WKB method for nonlinear equations.
I.  Linear theory,  Basel~-- Boston~-- Berlin, Birkhauser Verlag,
1994).

\bibitem{shapovalov:BEL-DOB} Belov V.V.,  Dobrokhotov S.Yu., Semiclassical
Maslov asymptotics with complex phases. I.  General approach,
{\it Teor. Mat. Fiz.}, 1992, V.92, N~2,  215--254 (English transl.: {\it  Theor. Math. Phys.},  1992, V.92, N~2,
843--868).

\bibitem{shapovalov:EHRENFEST}  Ehrenfest~P.,  Bemerkung \"uber die angenherte G\"ultigkeit
 der klassishen Mechanik innerhalb der Quanten Mechanik,
{\it Zeits. Phys.},  1927, Bd.45,  455--457.

\bibitem{shapovalov:MANKO}  Malkin~M.A.,  Manko~V.I.,
Dynamic symmetries and coherent states of quantum systems,
Nauka, Moscow, 1979 (in Russian).


\bibitem{shapovalov:PEREL} Perelomov~A.M.,  Generalized coherent states and their
application, Berlin, Springer-Verlag,  1986.



\bibitem{shapovalov:PUKHNACHEV}  Meirmanov~A.M., Pukhnachov~V.V.,  Shmarev~S.I.,
 Evolution equations and Lagrangian coordinates, New York~-- Berlin,
 Walter de Gruyter, 1994.


\end{thebibliography}
\end{document}